\documentclass[smallcondensed]{svjour3_edited}     
\smartqed  
\usepackage{graphicx}
\usepackage{amsmath, amssymb}
\usepackage{bm}
\usepackage{booktabs}
\usepackage{multirow}
\usepackage{float} 
\usepackage[usenames, dvipsnames]{color}
\usepackage{natbib}
\newcommand{\T}{\mathrm{\scriptscriptstyle T}}

\begin{document}
\title{KOALA: A new paradigm for election coverage 
}
\subtitle{An opinion poll based ``now-cast'' of probabilities of events in multi-party electoral systems}
\titlerunning{KOALA: Coalition analyses}   

\author{Alexander Bauer \and Andreas Bender \and Andr\'e Klima \and Helmut K\"{u}chenhoff}


\institute{A. Bauer \at
              Statistical Consulting Unit StaBLab, Department of Statistics, LMU Munich, Germany \\
              Tel.: +49-89-2180-3197 \\
              Fax: +49-89-2180-5308 \\
              \email{alexander.bauer@stat.uni-muenchen.de} \\
              ORCID: 0000-0003-3495-5131
           \and
           A. Bender \at
              Statistical Consulting Unit StaBLab, Department of Statistics, LMU Munich, Germany \\
              ORCID: 0000-0001-5628-8611
           \and
           A. Klima \at
              Statistical Consulting Unit StaBLab, Department of Statistics, LMU Munich, Germany
           \and
           H. K\"{u}chenhoff \at
              Statistical Consulting Unit StaBLab, Department of Statistics, LMU Munich, Germany
}


\maketitle

\begin{abstract}\ \\
Common election poll reporting is often misleading as sample uncertainty is
addressed insufficiently or  not covered at all. Furthermore, main interest
usually lies beyond the simple party shares. For a more comprehensive
opinion poll and election coverage, we propose shifting the focus
towards the reporting of survey-based probabilities for specific events of interest.
We present such an approach for multi-party electoral systems, focusing on
probabilities of coalition majorities. A Monte Carlo approach based on a Bayesian
Multinomial-Dirichlet model is used for estimation.
Probabilities are estimated, assuming the election was
held today (``now-cast''), not accounting for potential shifts in the electorate
until election day (``fore-cast''). Since our method is based on the posterior
distribution of party shares, the approach can be used to answer a variety of
questions related to the outcome of an election. We also introduce visualization
techniques that facilitate a more adequate depiction of relevant quantities as
well as respective uncertainties. The benefits of our approach are
discussed by application to the German federal elections in 2013 and 2017.
An open source implementation of our methods is freely available in the
\textbf{\texttt{R}} package \texttt{coalitions}.

\end{abstract}

\section{Introduction} \label{intro}
In multi-party democracies, approval of the government's  and the opposition
parties' work is usually measured by public opinion polls continuously conducted
and published by various polling agencies. Reported quantities
usually include the share of respondents that would vote for the respective
political parties {\it if the election was held today} (party shares),
the number of overall respondents and -- often less prominent -- information
about sample uncertainty.\\

One party often does not obtain enough votes for a governance majority on its own,
if the voting system has some kind of proportional allocation of seats in
parliament. Thus, multiple parties form
a so-called \emph{coalition} to jointly obtain the necessary majority of seats
in parliament. Media usually reports the results of opinion polls by focusing
on the reported party shares while ignoring sample uncertainty. This is misleading,
especially if shares are used to infer the possibility of a majority for
a specific coalition. For example, in the prelude to the 2017 German federal election,
a coalition was oftentimes stated to ``lose'' its majority just because the reported joint
voter share dropped below 50\% from one opinion poll to the next \citep[e.g.,][]{umfrage_2017}.
Such interpretations are clearly inadequate as sample uncertainty
(and often redistribution of votes) is not taken into account. This becomes especially
problematic, when one or more parties are close to the country specific threshold
of votes that has to be passed in order to enter the parliament. This was the case
in the 2013 German federal election, where the reported share of the Free Democratic
Party (FDP) was slightly above the 5\% threshold but failed to enter the
parliament on election night (cf. Section \ref{ssec:intro-ex-fdp}).\\


Beyond ensuring proper reporting of sample uncertainties, in our opinion, the
focus of election poll reporting should in general be shifted away from the
reported party shares. Instead, election coverage should focus on the most relevant
question, i.e., how {\it probable} is a specific event or outcome, given
the current political mood. As probabilities combine both, the reported shares and
sample uncertainty in one number, they allow more precise as well as more adequate
statements about specific events. Before an election, events such as the
following are usually of interest:

\begin{itemize}
  \item ``Will a party obtain enough votes to enter the parliament (pass the threshold)?''
  \item ``Will a party obtain the most (second most, third most, etc.) votes?''
  \item ``Will a specific coalition obtain enough votes (joint majority) to form a governing coalition?''
\end{itemize}

In this article, we present our approach for election and coalition analysis
(in German: \textbf{Koal}itions-\textbf{A}nalyse (KOALA)) that estimates probabilities
for any such events, referred to as \emph{p}robability \emph{o}f \emph{e}vent (POE)
in the following. In Section \ref{sec:application}, we will illustrate that the
POE brings more value to opinion poll based election coverage.
It is important to note that we quantify the contemporary political mood
and the resulting event probabilities (``now-cast''), not taking into consideration
potential shifts until election day (``fore-cast''). Approaches for predicting
future election outcomes based on past information can, e.g., be found in
\citet{graefe_2017} or \citet{norpoth_gschwend_2010}. A special focus is put on
multi-party proportional representation electoral systems and the estimation
of probabilities for (joint) majorities. POEs are estimated by Monte Carlo
simulations of election outcomes from the Bayesian posterior distribution of party
shares conditional on current observed opinion poll data. Prior to the German general
elections 2013 and 2017, results based on (an earlier iteration of) our approach
already entered media reporting \citep[cf.][]{wahlistik_2013, gelitz_2017}.\\

All methods discussed in this article are implemented in \texttt{R} \citep{r_2017}
and are available in the open-source package \texttt{coalitions} \citep{bender_bauer_2018}.
A \texttt{shiny}-based \citep{chang_2017} website
\texttt{koala.stat.uni-\allowbreak muenchen.\allowbreak de} visualizes estimated
coalition probabilities and is used to communicate the results for German federal
and state elections to the general public. Additionally, we applied our method
to the Austrian federal election 2017. The process of fetching new polls,
updating the website and sending out Twitter messages based on the newest results
is automated and allows for an immediate transfer of the estimated POE
to media outlets as well as the general public.

\subsection{Data basis}\label{ssec:data-basis}
As data base for our calculations, we use opinion polls conducted by established
polling agencies that quantify the electoral mood in a limited time frame
(\textit{if an election was held today}). For each of the two elections discussed
in Section \ref{sec:application}, we base the discussion on opinion polls published
by major German polling agencies (i.e., Allensbach, Emnid, Forsa, Forschungsgruppe Wahlen,
GMS, Infratest dimap and INSA), starting one year before each election.
Opinion poll data from these polling agencies is collected by and made publicly available
on \texttt{www.wahlrecht.de}. Application of our approach to other countries
requires systematic access to respective polling data. For the Austrian general
election in 2017 for example, we used the data base available at
\texttt{https://neuwal.com/wahlumfragen/}.

\subsection{Motivating example}\label{ssec:intro-ex-fdp}
In the last opinion poll conducted before the German federal election 2013 \citep{forsa_2013},
it was of special interest whether the conservative ``Union'' -- i.e., the union of
the parties CDU (Christian Democratic Union) and CSU (Christian Social Union in Bavaria ) --
and the liberal FDP would together once again obtain enough votes to form the
governing coalition (cf. Table \ref{tab_fdp}).

\begin{table}[!ht]\centering
\caption{Reported party shares in the Forsa opinion poll for the German federal
election, published September 20th, 2013 with $n=1995$ respondents.
\label{tab_fdp}
}
\medskip
\begin{tabular}{cccccccc}
\toprule[0.09 em]
Union & SPD & Greens & FDP & The Left & Pirates & AfD & Others \\
\midrule
40\% & 26\% & 10\% & 5\% & 9\% & 2\% & 4\% & 4\% \\
\bottomrule[0.09 em]
\end{tabular}
\end{table}

The German election system mandates a 5\% vote share threshold for parties to
enter the parliament. Votes for parties below this threshold and without at least
three successful direct candidates are redistributed (proportionally) to parties
above it. Table \ref{tab_fdp_redist} depicts the resulting redistributed party
shares given the poll in Table \ref{tab_fdp}. It illustrates, that Union-FDP with
its reported joint 45\% voter share before redistribution, would obtain 50\% of
parliament seats after redistribution. Thus, ignoring uncertainty it could be
concluded that a majority for this coalition is possible, if party shares would
increase slightly for one of the two parties.

\begin{table}[!ht]\centering
\caption{Redistributed party shares based on the Forsa opinion poll for the German
federal election, published September 20th, 2013 with $n=1995$ respondents
(cf. Table \ref{tab_fdp}). Party shares marked with ''--'' indicate parties that
would not pass the 5\% threshold (before redistribution).
\label{tab_fdp_redist}
}
\medskip
\begin{tabular}{cccccccc}
\toprule[0.09 em]
Union & SPD & Greens & FDP & The Left & Pirates & AfD & Others \\
\midrule
44.44\% & 28.89\% & 11.11\% & 5.56\% & 10.00\% & -- & -- & -- \\
\bottomrule[0.09 em]
\end{tabular}
\end{table}

However, such a consideration completely ignores sample uncertainty and the
probabilistic nature of the outcome. If the poll in Table \ref{tab_fdp} is
representative for the electoral mood, one would expect that the FDP enters the
parliament (passes the 5\% threshold) with a
probability of about $50\%$. Thus, the (posterior) distribution of the joint voter
share is bimodal and also depends on whether the other ``small'' parties close to
the 5\% threshold enter the parliament. The example also illustrates that discussion
of reported party shares can become very complex, due to sample uncertainty
and the multitude of different outcomes this uncertainty entails.
We therefore argue that probability based reporting of opinion poll results can
answer the actual question of interest (``Will a coalition of Union-FDP obtain
enough votes to obtain a majority of seats in the parliament?'') more directly, while
adequately taking into account the inherent uncertainty.\\

The remainder of the article is structured as follows: Section \ref{sec:methods}
introduces the Bayesian method used to estimate POEs as well
as some details on the aggregation of multiple opinion polls and the correction
of rounding errors. Section \ref{sec:application} illustrates the
application of the approach to opinion polls in advance of the  2013 and 2017
German federal elections. A summary and discussion are presented in Section
\ref{sec:conclusion}.\\

\section{Methods}\label{sec:methods}

\subsection{Estimating event probabilities from reported party shares} \label{ssec:bayes}
To estimate the POE conditional on opinion poll results we use the Bayesian
framework to construct the \emph{posterior} distribution of the party shares based
on distribution of the reported shares and an assumption
about their \emph{prior} distribution.\\

Let $X_1,\ldots, X_P$ be the reported opinion poll count of respondents that would
elect party $p, \ p=1,\ldots,P$ (vote count). For example, in Table \ref{tab_fdp}
the reported vote count for the \emph{Union} is given by $X_1 = .40 \cdot 1995 = 798$.
We assume that $(X_1, \ldots, X_P)^\T$ follows a Multinomial distribution
\begin{equation}\label{eq:multinom}
\boldsymbol{X} = (X_1,\ldots, X_P)^\T \sim Multinomial(n, \theta_1,\ldots, \theta_P),
\end{equation}

where $n$ is the sample size of the opinion poll and $\theta_p, \ p=1,\ldots,P$
indicates the probability of party $p$ being selected. Further assuming a
simple random sample, i.e ignoring possible bias,
$\theta_p$ represents the true percentage of voters for party $p$ in
the general population.
 Given one (pooled) survey,
the distribution of the observed vote counts $\mathbf{x}=(x_1,\ldots,x_P)^\T$ is
denoted by $f(\mathbf{x}|\boldsymbol{\theta})$.\\

For the prior distribution of the true party shares $\boldsymbol{\theta}=(\theta_1,\ldots, \theta_P)^\T$
we chose an uninformative prior distribution
(Jeffrey's prior; \citet{gelman_2013})

\begin{equation}\label{eq:prior}
\begin{aligned}
\boldsymbol{\theta} &= (\theta_1,\ldots,\theta_P)^\T \sim Dirichlet(\alpha_1,\ldots,\alpha_P), \\
\text{with} &\ \ \ \ \ \ \ \ \ \ \ \ \ \ \ \alpha_1 = \ldots = \alpha_P = \frac{1}{2},
\end{aligned}
\end{equation}

denoted by $p(\boldsymbol{\theta}|\boldsymbol{\alpha})$.
As the Dirichlet distribution is a conjugate prior to the Multinomial distribution,
the resulting posterior distribution \eqref{eq:posterior-dens}  of parameters $\boldsymbol{\theta}|\mathbf{x}$

\begin{align}
f(\boldsymbol{\theta}|\mathbf{x})
  & = \frac{f(\mathbf{x}, \boldsymbol{\theta})}{f(\mathbf{x})}
    = \frac{f(\mathbf{x}|\boldsymbol{\theta})p(\boldsymbol{\theta}, \boldsymbol{\alpha})}{f(\mathbf{x})}\label{eq:posterior-dens}\\
  & \propto f(\mathbf{x}|\boldsymbol{\theta})p(\boldsymbol{\theta}|\boldsymbol{\alpha})\\
  & \propto \prod_{p=1}^{P}\theta_p^{x_p}\cdot \prod_{p=1}^{P}\theta_p^{\alpha_{p}-1} = \prod_{p=1}^{P}\theta_p^{x_p + \alpha_p-1},
\end{align}

is again a Dirichlet distribution with
\begin{equation}\label{eq:posterior}
\boldsymbol{\theta}|\mathbf{x} \sim Dirichlet(x_1 + 1/2,\ldots, x_P + 1/2).
\end{equation}

Given the multivariate posterior \eqref{eq:posterior} and using Monte Carlo
simulations, POEs can be deduced for many types of events by simulating
election results from \eqref{eq:posterior} and calculating the percentage of
simulations in which the event of interest occurred. This includes the
probabilities for specific majorities derived from a complex, country-specific
system of rules for the calculation of seats in the parliament
(Sainte-Lague/Scheppers in Germany; \citet{grofman_2003}). For example, given the
Forsa poll introduced in Section \ref{ssec:intro-ex-fdp}, the coalition of Union-FDP
obtained a majority of seats in $2\,633$ of $10\,000$ simulations, which equals
an POE of $26\%$ (see Section \ref{sec:application} for more details).\\

If it is known that estimates of specific party shares are biased for some opinion
polls/agencies, this information could be included in the model by using an
informative prior distribution. The prior parameters $\alpha_p$ would then
be adjusted to have higher or lower values, respectively.
However, such biases of polling agencies are hard to quantify as the true party
share in the electorate is only known on election days. For our analyses, we
therefore use the uninformative prior \eqref{eq:prior}.

\subsection{Aggregation of multiple polls (Pooling)} \label{ssec:pooling}
In the presence of multiple published opinion polls, pooling is used to
aggregate multiple polls in order to reduce sample uncertainty. To ensure a
reliable pooling regarding the current public opinion, we only use polls
published within a certain period of time (e.g. 14 days)  and only use the most recent survey published by each polling agency.\\

Considering a single poll $i$, the observed vote count $X_{ip}$ for each of
$P$ parties follows a multinomial distribution with sample size $n_i$ and underlying,
unknown party shares $\theta_p$ in the population.
Pooling over multiple such polls as independent random samples leads to another
multinomial distribution for the summed number of votes $\sum_i X_{ip}$:
\begin{equation}
\left( \sum\limits_i X_{i1},\ldots, \sum\limits_i X_{iP} \right)^\T
  \sim Multinomial \left( \sum\limits_i n_i,\theta_1,\ldots,\theta_P\right).
\end{equation}

Further analyses, however, showed that polls from different (German) polling
agencies are correlated and the independence assumption does not hold. Therefore,
we adjust the resulting multinomial distribution by using an
\textit{effective sample size} \citep{hanley_2003}, reflecting that the aggregation
over multiple correlated polls does not contain information of a sample with
$\sum_i n_i$ observations.\\

Quantification of pairwise correlation is done based on the variance of the party
share difference between two polls for a specific party. The following
holds for two independent random samples from poll $A$ and $B$:

\begin{equation}
\begin{aligned}
Var(X_{Ap} - X_{Bp}) &= Var(X_{Ap}) + Var(X_{Bp}) - 2 \cdot Cov(X_{Ap}, X_{Bp}) \\
\Leftrightarrow \ \ \ \ Cov(X_{Ap}, X_{Bp}) &= \frac{1}{2} \cdot \left(Var(X_{Ap}) + Var(X_{Bp}) - Var(X_{Ap} - X_{Bp}) \right).
\end{aligned}
\end{equation}

We take $Var(X_{Ap})$ and $Var(X_{Bp})$ as the theoretical variances of the
binomially distributed, reported vote count and estimate $Var(X_{Ap} - X_{Bp})$
based on the observed differences between the reported party shares. Having done so,
it is possible to estimate the covariance $Cov(X_{Ap}, X_{Bp})$ and accordingly
also the correlation. As the binomial variance is directly proportional to sample
size, the effective sample size $n_{\text{eff}}$ can be defined as the ratio
between the estimated variance of the pooled sample and the theoretical variance
of a sample of size one:
$$
n_{\text{eff}} = \frac{Var(\text{pooled})}{Var(\text{sample of size 1})}.
$$
In the case of two surveys,
$$
Var(\text{pooled}) = Var(X_{Ap} + X_{Bp}) = Var(X_{Ap}) + Var(X_{Bp}) + 2 \cdot Cov(X_{Ap},X_{Bp})
$$
and $Var(\text{sample of size 1})$ the theoretical variance of the pooled share.

Considering the party-specific correlations between 20 surveys conducted by the
two German polling agencies that provide updates most regularly, Emnid and Forsa,
we on average end up with a medium high positive correlation, using mean party
shares and sample sizes per institute for the theoretical variances. Comparisons
of other agencies were not performed as too few published surveys that
cover comparable time frames were available. For simplicity, we do not recalculate
the correlation for each simulation, but rather set the correlation used in our
calculations to $0.5$, i.e., a medium positive correlation. For convenience, the
calculation of $n_{\text{eff}}$ is based on the party with most votes, as the
specific party choice only marginally affects the results.\\

Considering for example two polls with $1\,500$ and $2\,000$ respondents,
respectively, and a pooled share of $40\%$ for the strongest party, the method
leads to an effective sample size of $n_{\text{eff}} = 2\,341$. Thus, the method
reduces sample uncertainty compared to using a single poll, while being quite
conservative compared to the assumption of independence which would lead to
an aggregate sample size of $1\,500 + 2\,000 = 3\,500$.\\

As noted above, in practice we use a time window of 14 days, i.e., all surveys
published in the last 14 days are included in the calculation of the pooled
sample. For some elections (e.g., state elections), opinion polls are
updated very rarely. In such cases the time window and pooling procedure
could be further modified, e.g., by including all surveys published within 14 days with full
weight (using their reported sample size), and all surveys that were published
between 15 and 28 days ago with halved weight (using the halved sample size).

\subsection{Correction of rounding errors}\label{ssec:rounding}
Polling agencies usually only publish rounded party shares and
raw data is not available. Therefore, we adjust the reported data by adding
uniformly distributed random noise to the observed party shares $\tilde{\theta}_p$ in order
to avoid potential biases caused by the use of rounded numbers:

\begin{equation}
\begin{aligned}
\tilde{\theta}_{p,adj} = \ &\tilde{\theta}_p + r_{\gamma,p}, \\
\text{with} \ \ \ \ \ &r_{\gamma,p} \sim U[-\gamma,\gamma].
\end{aligned}
\end{equation}

The correction coefficient $\gamma$ is chosen according to rounding accuracy.
E.g., for data rounded to $1\%$ steps we use $\gamma = 0.5\%$. After random noise
was added the adjusted shares are rescaled to ensure that all adjusted party shares
$\tilde{\theta}_{p,adj}$ sum to $100\%$. Overall, instead of using rounded numbers and simulating
$n_s$ values from the resulting posterior, we perform $n_s$ simulations where we
first adjust the party shares using individually drawn $r_{\gamma,p}$ and
then simulate one observation from each resulting posterior.\\

\section{Application} \label{sec:application}
An earlier iteration of our method entered media reporting before the German
federal elections 2013 and 2017 \citep[cf.][]{wahlistik_2013, gelitz_2017}.
We will discuss these two elections in order to elaborate the differences between
standard media coverage of election polls -- focused on the interpretation of the
reported party shares -- and our approach based on estimated POEs. Reported party
shares as described in Section \ref{ssec:data-basis} were used as data basis.
Polls from different agencies published within a time window of 14 days were
aggregated (cf. Section \ref{ssec:pooling}). For the estimation of POEs
$n_{s} = 10\,000$ simulations were performed.

\subsection{German federal election 2013} \label{subsec:2013}
\begin{figure}[H]\centering
\includegraphics[width=0.6\textwidth]{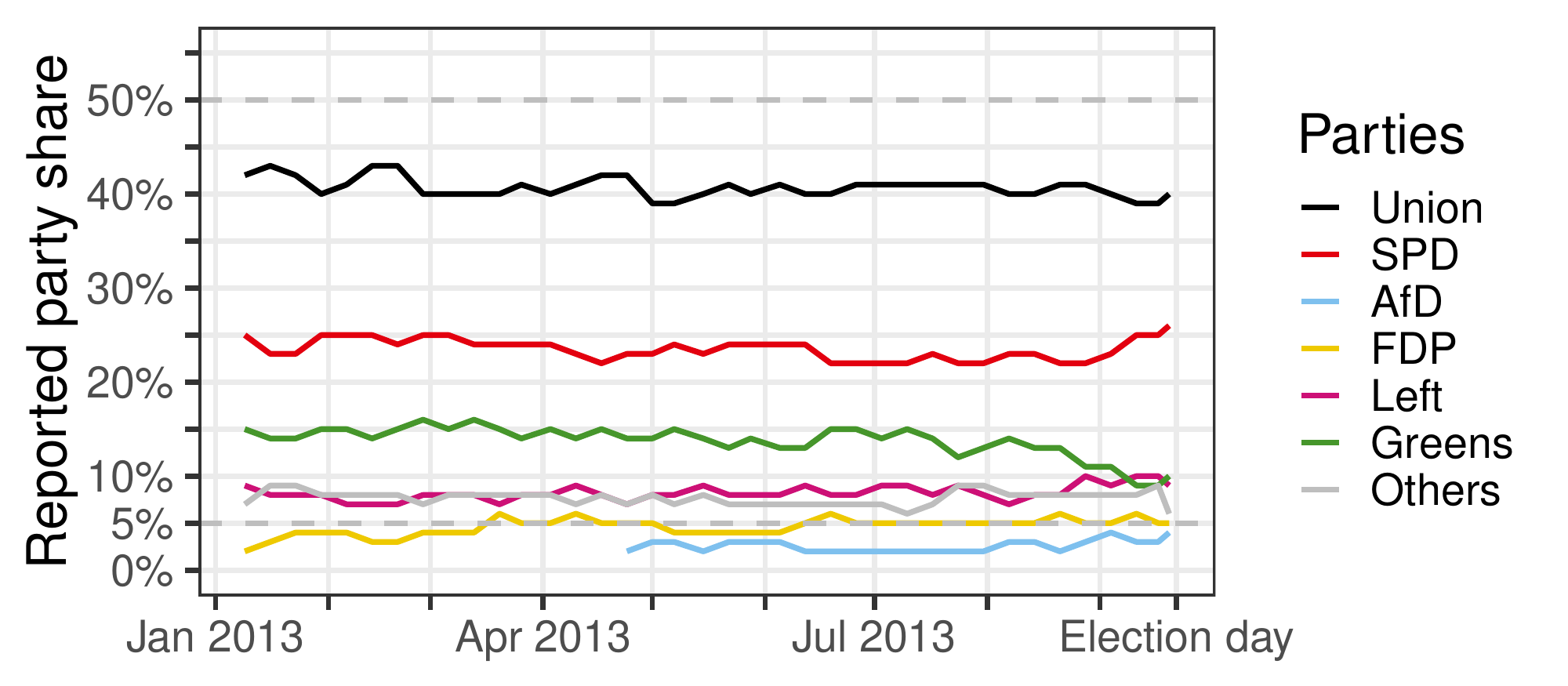}
\caption{Reported party shares based on Forsa opinion polls 
from October 2012 until election day on September 22nd, 2013.
Shares for the AfD were only explicitly reported starting in April 2013.
Before that time, the party is contained in ``Others''.
\label{fig:2013}
}
\end{figure}

In the legislative period from 2009 to 2013, the German government was formed by
a coalition of the conservatives (Union) and the liberals (FDP). Before the
election on September 22nd, 2013, the question whether the coalition could
sustain its majority was therefore of main interest. The FDP played a key role,
as the coalition could only be formed if the party had reached the minimum share
of $5\%$ of the votes. Figure \ref{fig:2013} summarizes the reported party shares
for the one year period prior to the election.

\paragraph{POE: FDP passing the 5\% threshold} \ \\

\begin{figure}[H]\centering
\begin{tabular}{ll}
\includegraphics[height=.2\textwidth]{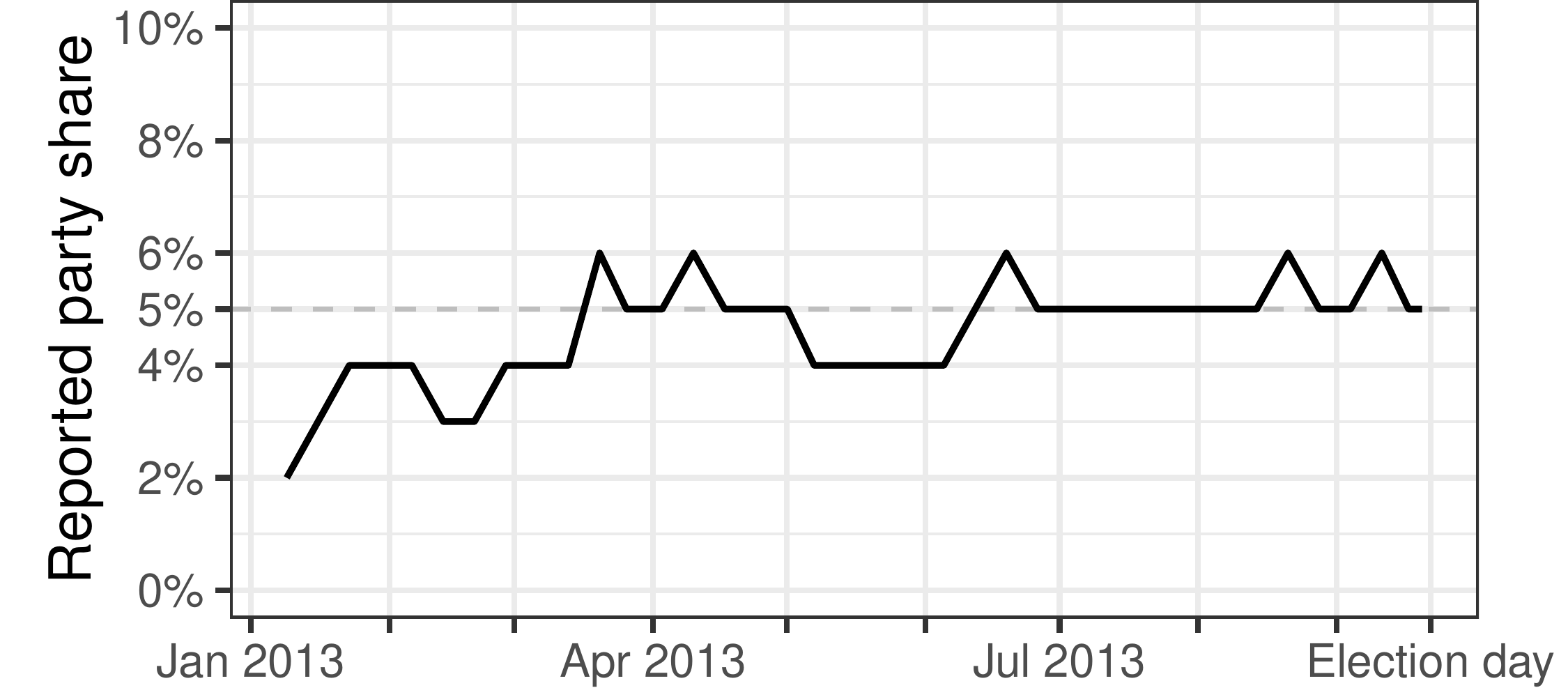}
&
\multirow{2}{*}[18.3ex]{\includegraphics[height=40ex]{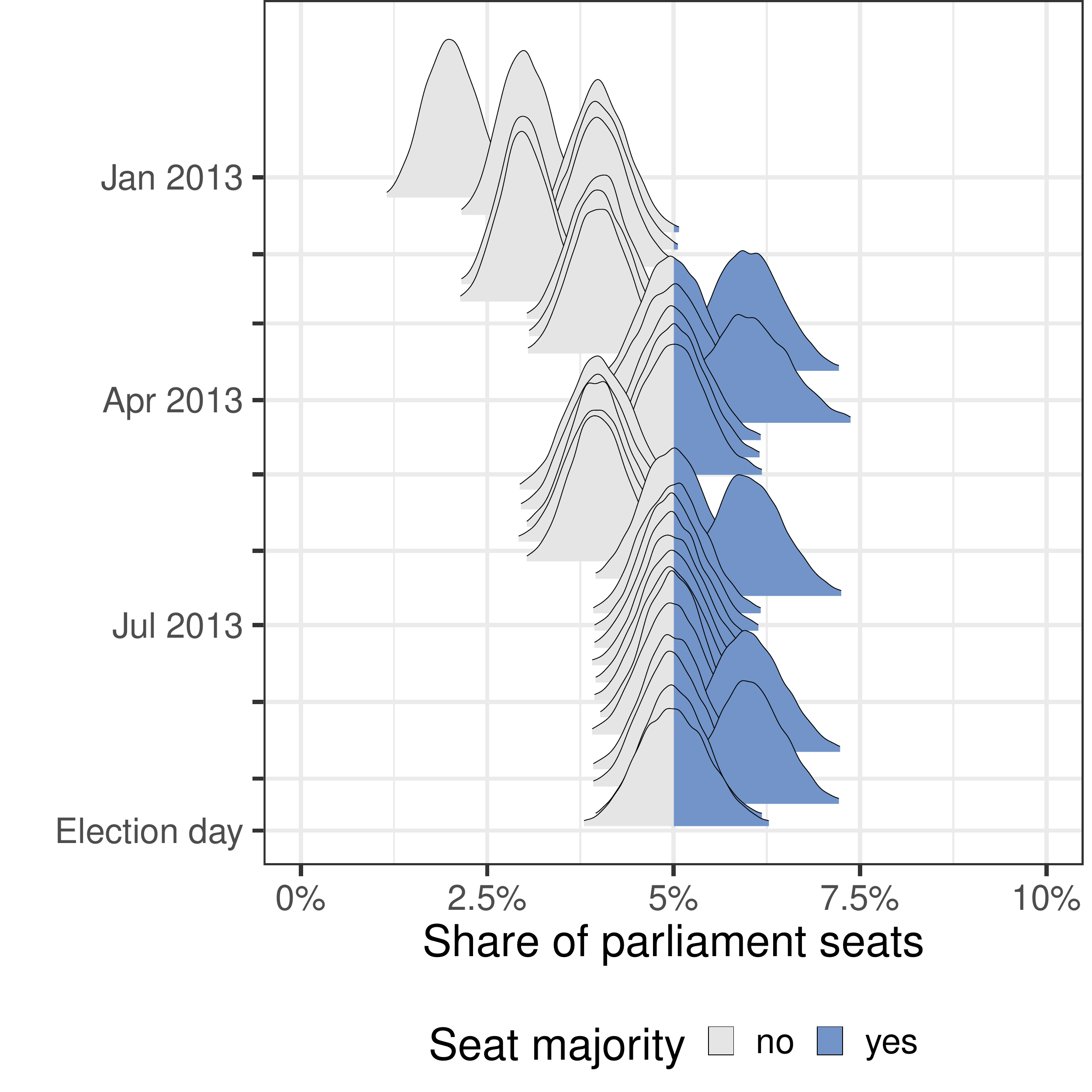}}
\\
\includegraphics[height=.2\textwidth]{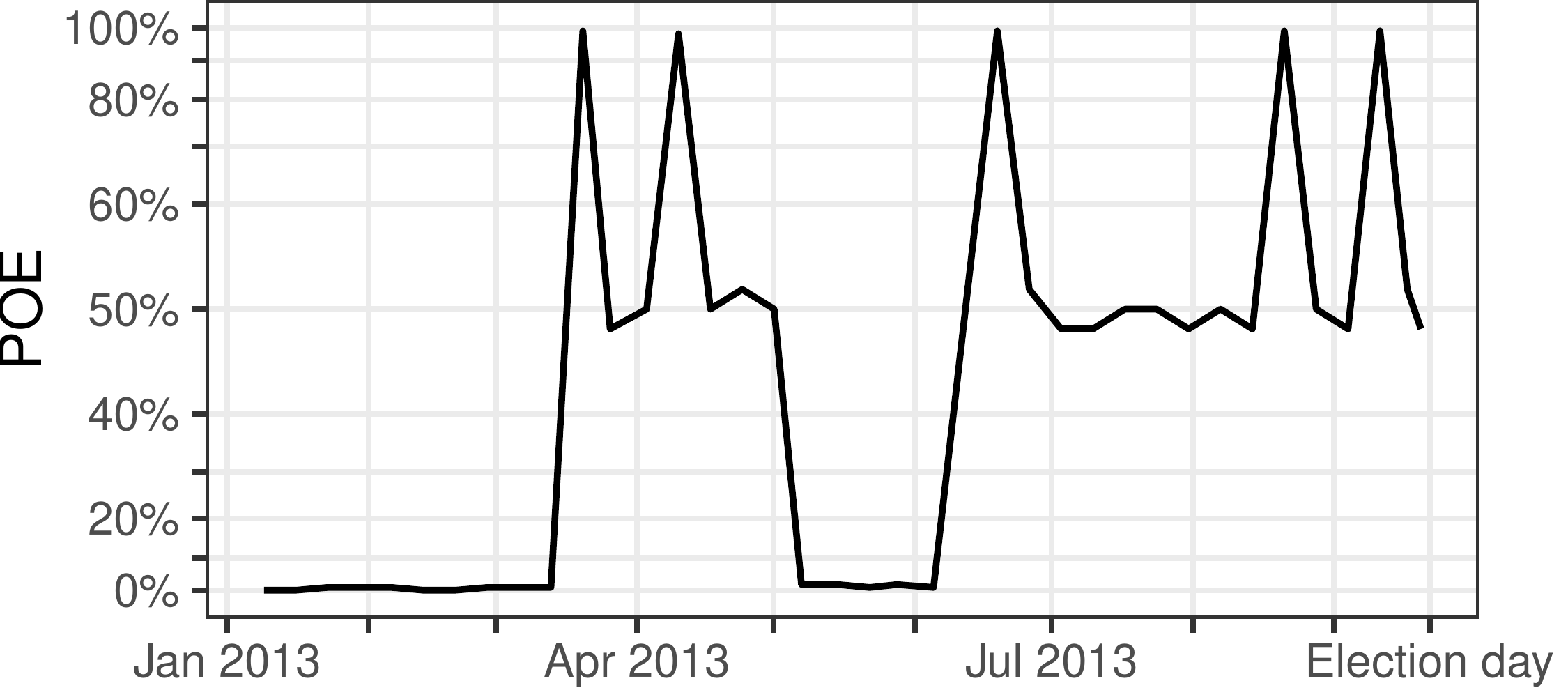}
\end{tabular}
\caption{Prospect of FDP to pass the $5\%$ threshold before the
German federal election in September 2013 based on Forsa opinion polls.
Top left: Reported party share before redistribution. Bottom Left: Now-cast
of the POE that FDP will pass the $5\%$ threshold, based on $10\,000$ simulations.
Right: Densities of the $10\,000$ simulated FDP shares. Areas under
the density colored blue indicate the simulations in which the FDP passes the
5\% threshold.
\label{fig:2013_fdp}
}
\end{figure}

The poll-based prospect of FDP to successfully pass into parliament is visualized
in Figure~\ref{fig:2013_fdp}.
As can be seen, the reported party share clearly exceeded the necessary hurdle
of $5\%$ only over short periods of time with maximum values
of $6\%$ (top left pane in Fig.~\ref{fig:2013_fdp}). Similarly, the now-cast POE
for the party to pass the threshold rarely rose over $50\%$ (bottom left pane).
In the last Forsa poll before election day, a party share of $5\%$ was reported,
stating that the event of FDP successfully passing into parliament was highly uncertain.\\

Comparing party shares and POEs, Figure~\ref{fig:2013_fdp} shows
that relatively small changes in the reported party share can dramatically influence
corresponding POE values, depending on the base level of the party share and
-- in this example -- its closeness to the $5\%$ threshold.
In this regard, probabilities make it easier to deduce {\it relevant} information
from opinion polls as they incorporate both the closeness of the reported shares
to the relevant threshold as well as sample uncertainty.
For example, party shares of $4\%$ and $6\%$
correspond to very definite POEs of near $0\%$ and $100\%$,
respectively and the reporting of such POEs leads to a much clearer perception
of the current public opinion compared to the reported party share and
survey sample size only. \\

\begin{figure}[H]\centering
\includegraphics[width=0.45\textwidth]{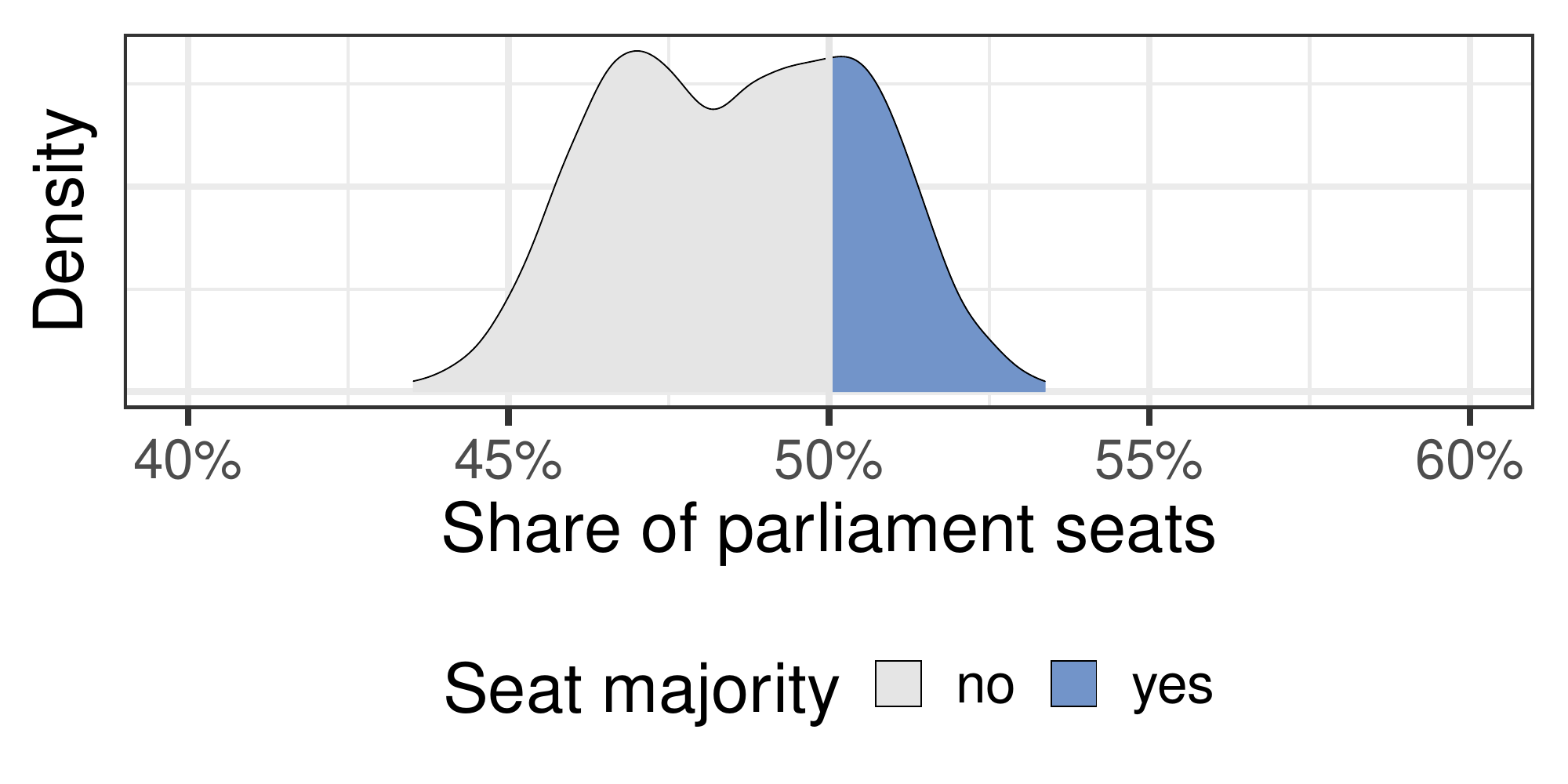}
\caption{Density of $10\,000$ simulated parliament seat shares for the coalition
Union-FDP before the German federal election in September 2013 based on the Forsa
opinion poll in Table \ref{tab_fdp}. The area under the density colored blue
indicates simulations with a Union-FDP majority, resulting in a
POE of about $26\%$.
\label{fig:seatDist}
}
\end{figure}

For the visualization of POEs we suggest to plot the distribution of
simulated shares via density plots and to highlight the area associated
with simulations where the event of interest occurred
(see also Fig. \ref{fig:seatDist}). This has the advantage that POEs are
communicated clearly (and intuitively) while the distribution of simulated
shares additionally highlights the underlying uncertainty and the range
of possible outcomes. Another advantage is that such visualizations can easily
be extended to depict the development of POEs over time using so-called
\emph{ridgeline plots} \citep{wilke_2017}. In Figure~\ref{fig:2013_fdp} this
plot type and the development of the POEs is compared to the observed 
FDP party shares usually reported in media. \\

To focus on the most relevant changes in the POEs, we propose
the use of a skewed axis as shown in Figure~\ref{fig:2013_fdp}. In this axis
the range of values around $50\%$ is stretched and the range of values near
$0\%$ and $100\%$ is compressed. In this way, we put less weight on changes
where an event is {\it still highly (im)probable} and emphasize more relevant
changes after which an event gets more or less probable than $50\%$, respectively. Also,
consistently using another axis for the estimated probabilities prevents
confusion of POEs and voter shares.

\paragraph{POE: Union-FDP coalition majority} \ \\

Figure \ref{fig:seatDist} shows the simulated parliament seat shares for the
coalition Union-FDP, based on the reported party shares in Table \ref{tab_fdp}.
The estimated density is clearly bimodal as the reported FDP share before
redistribution equals exactly $5\%$ and therefore FDP only enters the parliament
in about $50\%$ of the simulations. In this case, a majority was observed in about
one quarter of the simulations, leading to an estimated POE of $26\%$. \\

\begin{figure}[H]\centering
\begin{tabular}{ll}
\includegraphics[height=.2\textwidth]{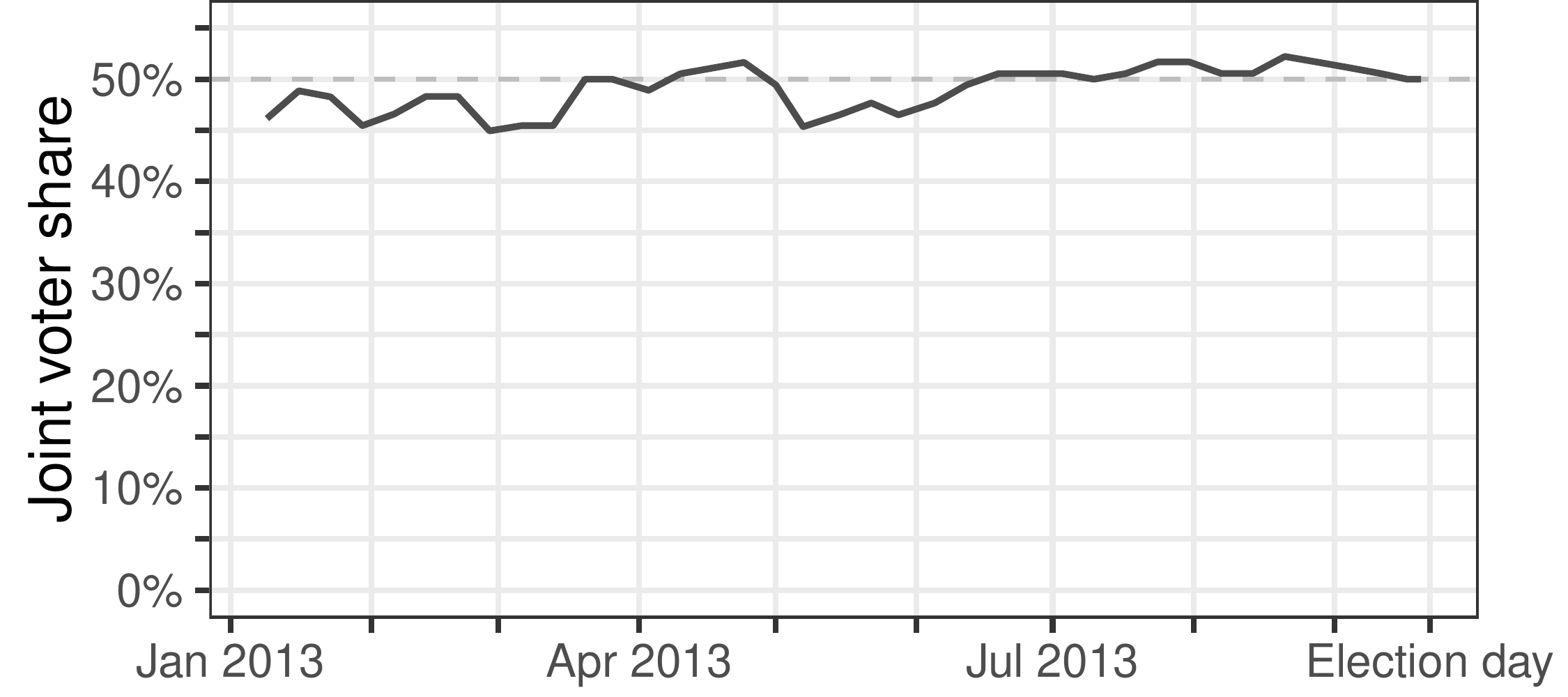}
&
\multirow{2}{*}[18.3ex]{\includegraphics[height=40ex]{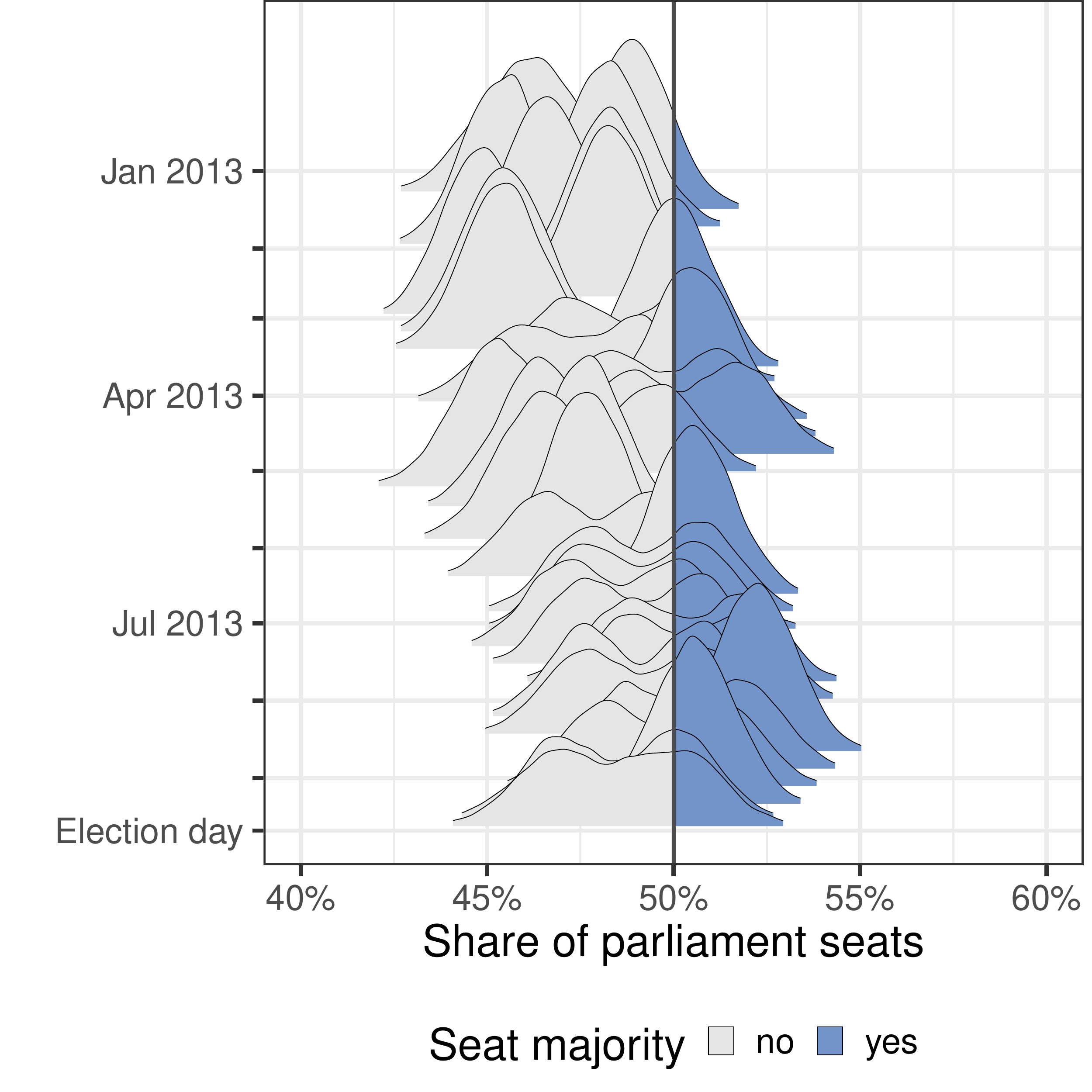}}
\\
\includegraphics[height=.2\textwidth]{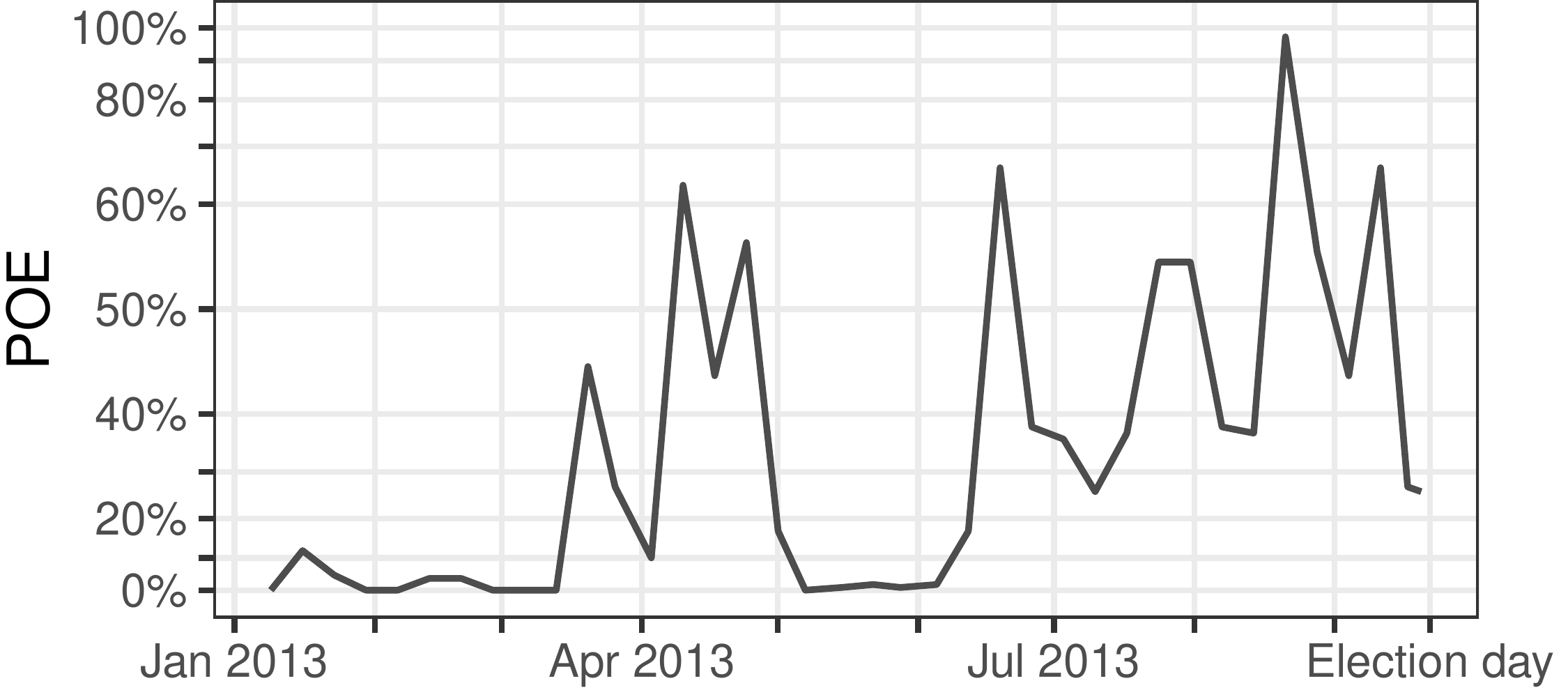}
\end{tabular}
\caption{Prospect of the coalition Union-FDP to obtain a government majority before the
German federal election in September 2013 based on Forsa opinion polls.
Top left: Reported joint voter shares after redistribution.
Bottom left: Now-cast of the POE that the coalition will obtain a government
majority, based on $10\,000$ simulations.
Right: Densities of the $10\,000$ simulated parliament seat shares. Areas under
the density colored blue indicate the simulations in which the coalition
obtains a parliament seat majority.
\label{fig:seatDist_time}
}
\end{figure}

Comparing the redistributed party shares and the POEs
in Figure~\ref{fig:seatDist_time}, it is again evident that even small changes
in the joint redistributed voter share can make an immense difference regarding the
POEs. Especially in the months preceding the election this leads to heavily
fluctuating POEs based on the Forsa opinion polls.
Furthermore, the development of the joint voter shares and the corresponding POEs
nicely highlights another advantage of using such probabilities. The POEs do not
only take into account the voter shares and sample uncertainty, but also
implicitly cover the uncertainty regarding whether FDP passess the $5\%$ threshold
or not. Between the middle of June and the middle of July the POE drops from 
nearly $70\%$ to under $30\%$, even though the joint voter share of Union-FDP
only changes marginally. Taking into account the development of reported FDP
party shares in Figure~\ref{fig:2013_fdp}, it becomes clear that this
drop is caused by a growing uncertainty of FDP passing into parliament.
As the FDP share drops to $5\%$ at the end of June also the POE for the
Union-FDP seat majority declines heavily. Accordingly, the densities in the
ridgeline plot in Figure~\ref{fig:seatDist_time} are unimodal or bimodal if the FDP
share is clearly above/below or close to the $5\%$ threshold, respectively. \\

\subsection{German federal election 2017} \label{subsec:2017}
\begin{figure}[H]\centering
\includegraphics[width=0.6\textwidth]{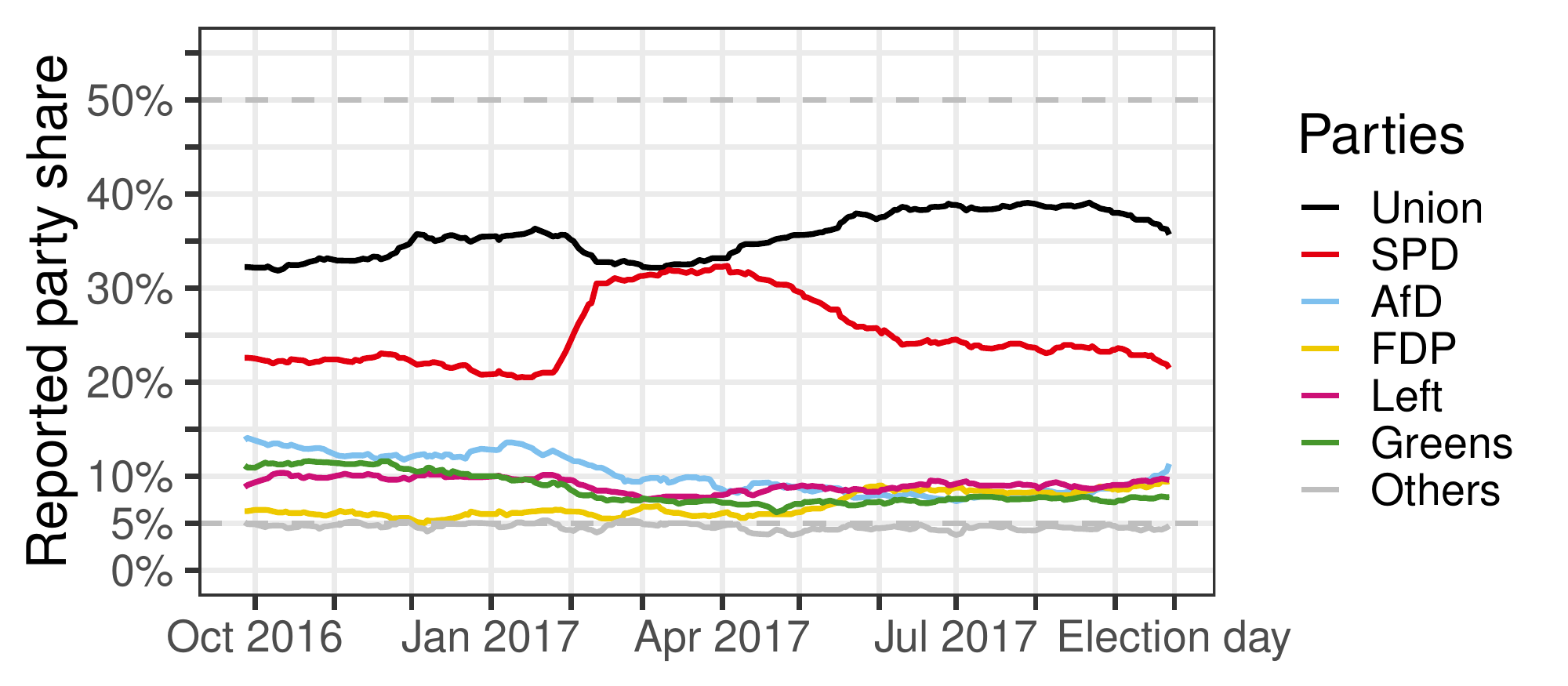}
\caption{Development of the pooled party shares from October 2016 until
election day on September 24, 2017, based on a pooling time window of 14 days.
\label{fig:2017}
}
\end{figure}

After the German federal election in 2013, a ``grand coalition'' between Union and the social
democratic SPD formed the government from 2013 to 2017.
For the following election on September 24th, 2017, the goal of both Union and SPD
was to obtain enough votes to form a coalition outside the grand
coalition. Therefore, multiple potential coalitions were of interest before the
election. In the following paragraphs, we will focus on the most prominently
discussed coalitions, i.e., the Union-led coalition Union-FDP, and the
SPD-led coalition of SPD, the Left party (Die LINKE) and the Green party, which
-- based on the joint voter share -- was the strongest alternative to a Union-led
government and was not clearly denied by the potential member parties until several weeks
before election day.
A question of major interest also was whether the right-wing party AfD, which
slightly missed the $5\%$ threshold in 2013, but gained support before the 2017 election,
would become the third strongest party in parliament after Union and SPD.
The pooled party shares before the 2017 election are shown in Figure~\ref{fig:2017}.

\begin{figure}[H]\centering
\begin{tabular}{ll}
\includegraphics[height=.2\textwidth]{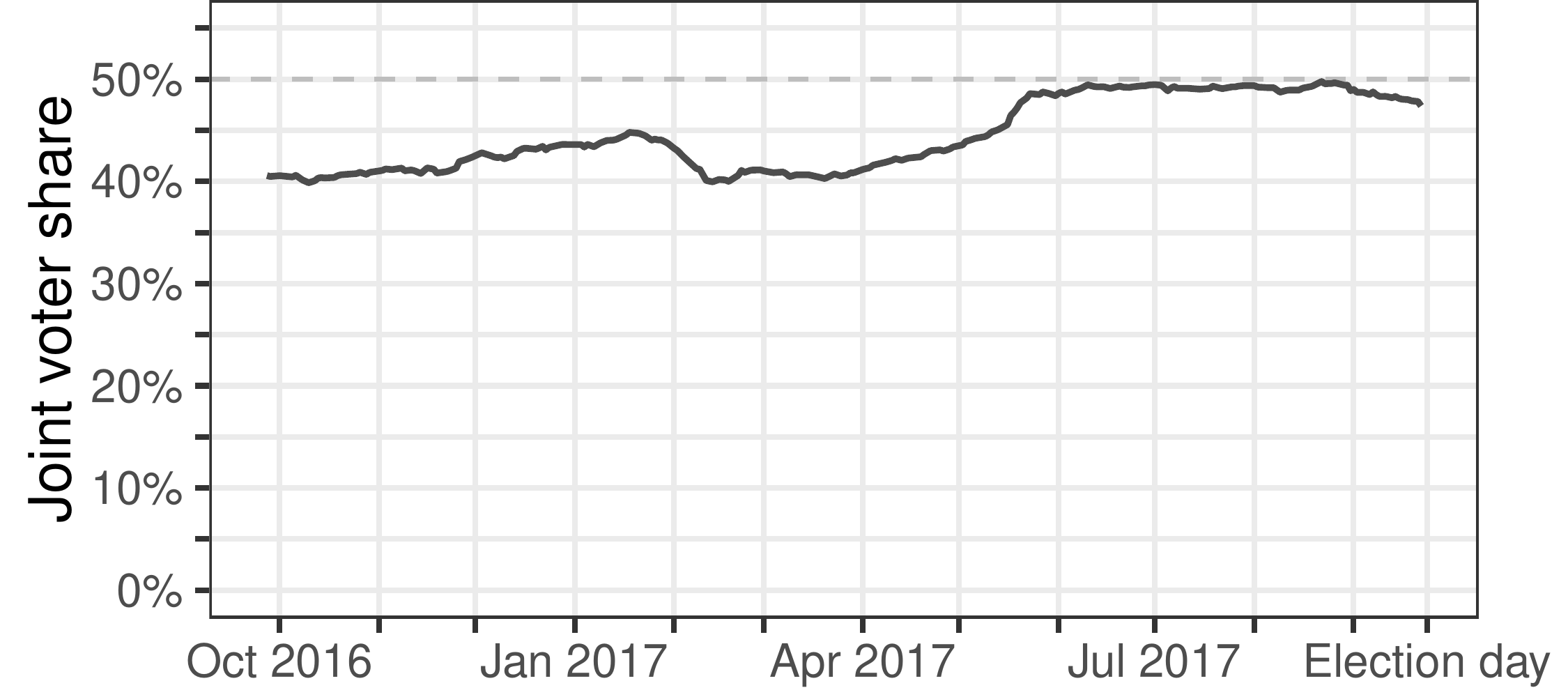}
&
\multirow{2}{*}[18.3ex]{\includegraphics[height=40ex]{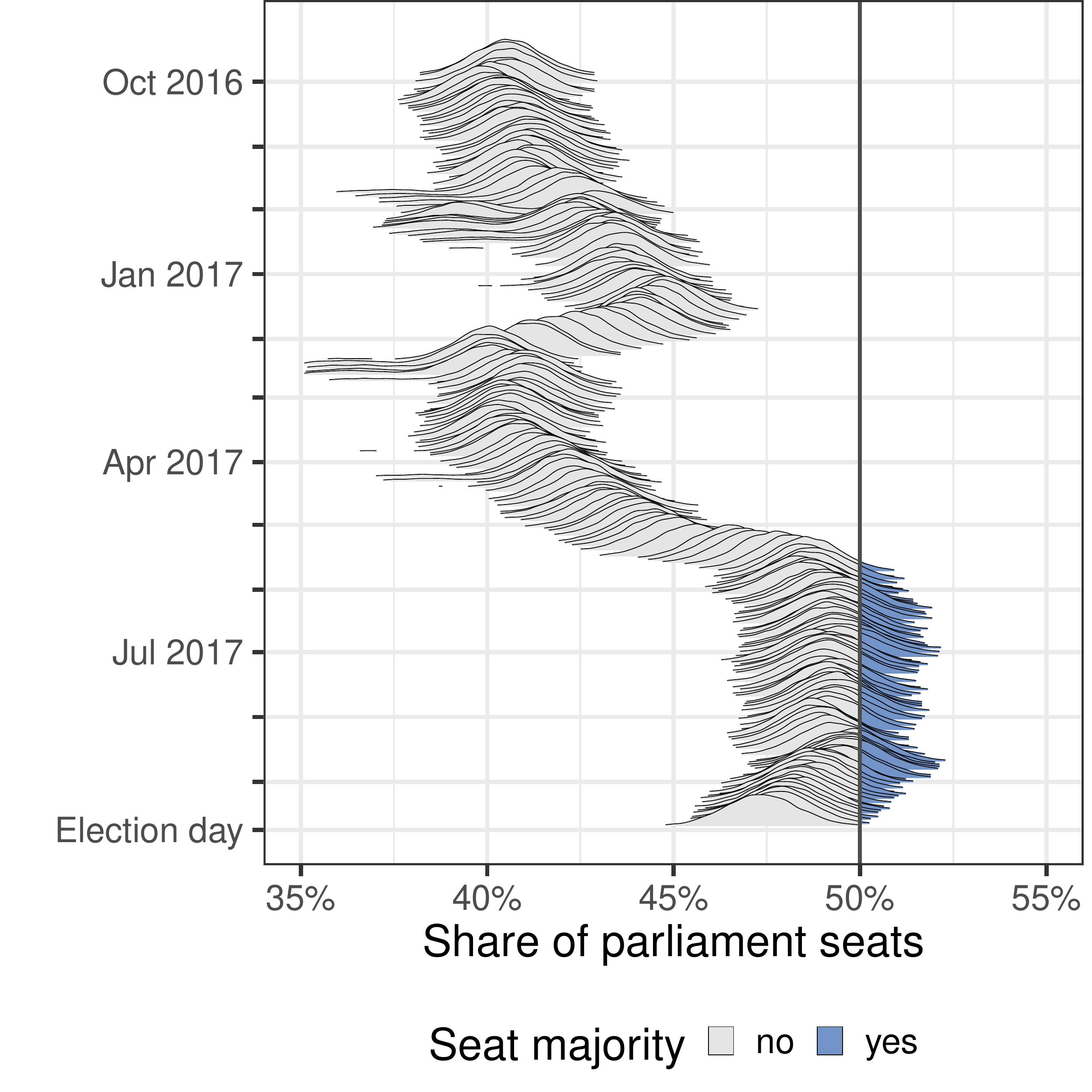}}
\\
\includegraphics[height=.2\textwidth]{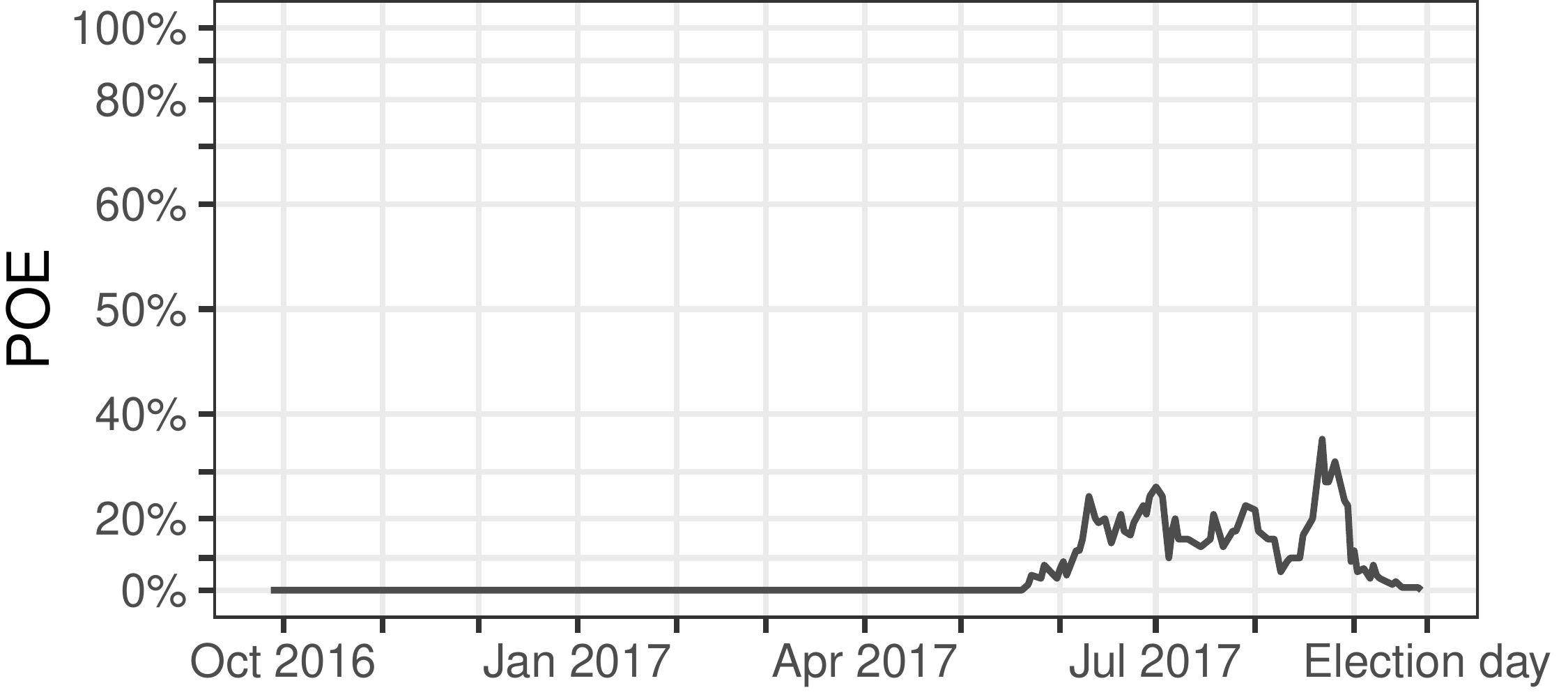}
\end{tabular}
\caption{Prospect of the coalition Union-FDP to obtain a government majority before the
German federal election in September 2017 based on pooled opinion polls.
Top left: Reported joint voter shares after redistribution.
Bottom left: Now-cast of the POE that the coalition will obtain a government
majority, based on $10\,000$ simulations.
Right: Densities of the $10\,000$ simulated parliament seat shares. Areas under
the density colored blue indicate the simulations in which the coalition
obtains a parliament seat majority.
\label{fig:2017_cdufdp}
}
\end{figure}

\paragraph{POE: Union-FDP coalition majority} \ \\
Compared to the German federal election in 2013, the situation for a
coalition between Union and FDP before the election in 2017
was quite different as FDP party shares were
clearly above the $5\%$ threshold (see Fig.~\ref{fig:2017}) most of the time.
However, as the share of Union was lower than in 2013,
the joint redistributed voter share was mostly below $50\%$.
As can be seen in Figure~\ref{fig:2017_cdufdp}, the coalition had
a joint, redistributed share of about $40\%$ in October 2016
and reached its maximum share of nearly $49.8\%$ about one month
before election day.

By comparison, the ridgeline plot in Figure~\ref{fig:2017_cdufdp}
shows that joint voter shares below $48\%$ correspond to very
small POEs of $<1\%$, based on pooled effective sample sizes of around $3\,000$.
On the other hand, based on comparable sample sizes, shares of $49\%$
and $49.5\%$ corresponded to probabilities of around $14\%$ and $25\%$,
respectively.
Overall, one month before election day the coalition had a good prospect
reaching a seat majority based on a redistributed share of $49.8\%$ and
a POE of nearly $40\%$. However, until two days before the election the pooled
share and the POE dropped to $47.4\%$ and $0.4\%$, respectively, making
a success of the two parties highly improbable.

\begin{figure}[H]\centering
\begin{tabular}{ll}
\includegraphics[height=.2\textwidth]{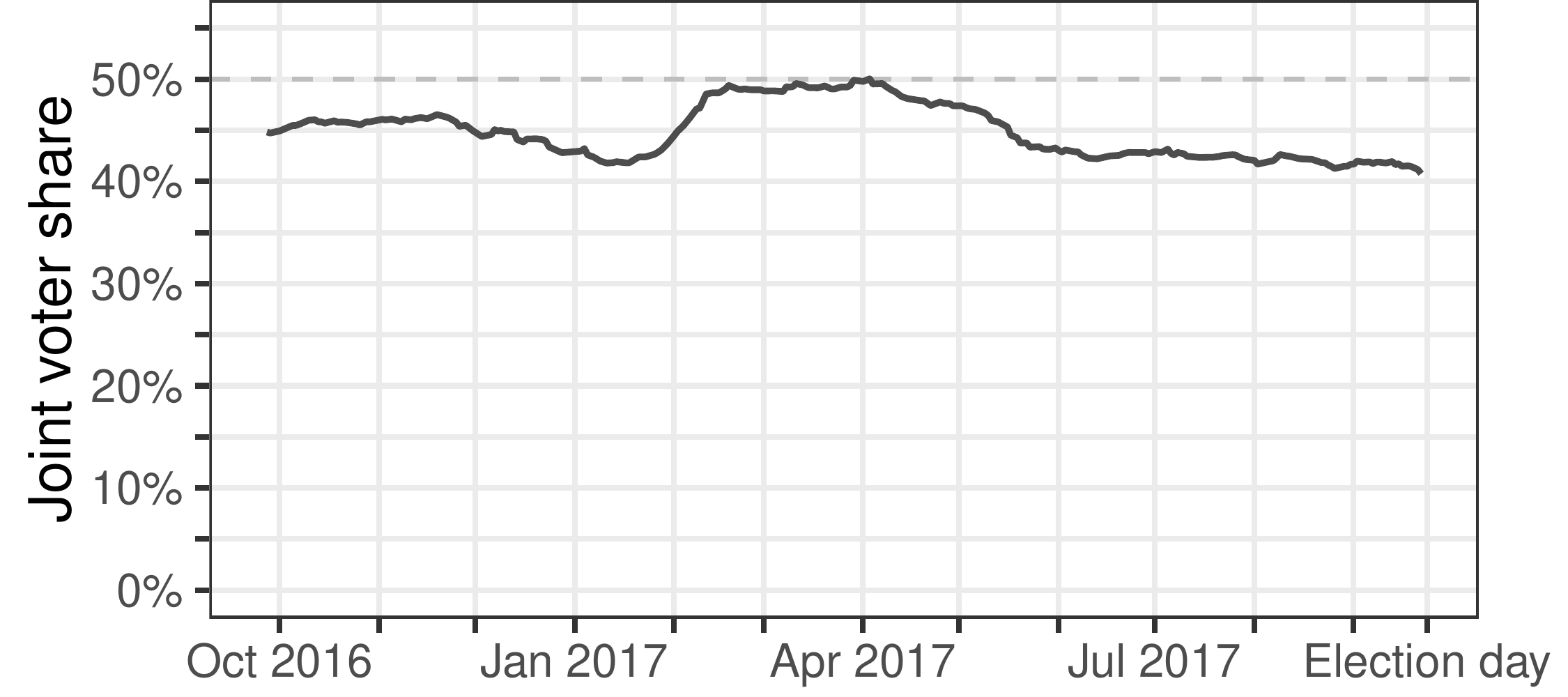}
&
\multirow{2}{*}[18.3ex]{\includegraphics[height=40ex]{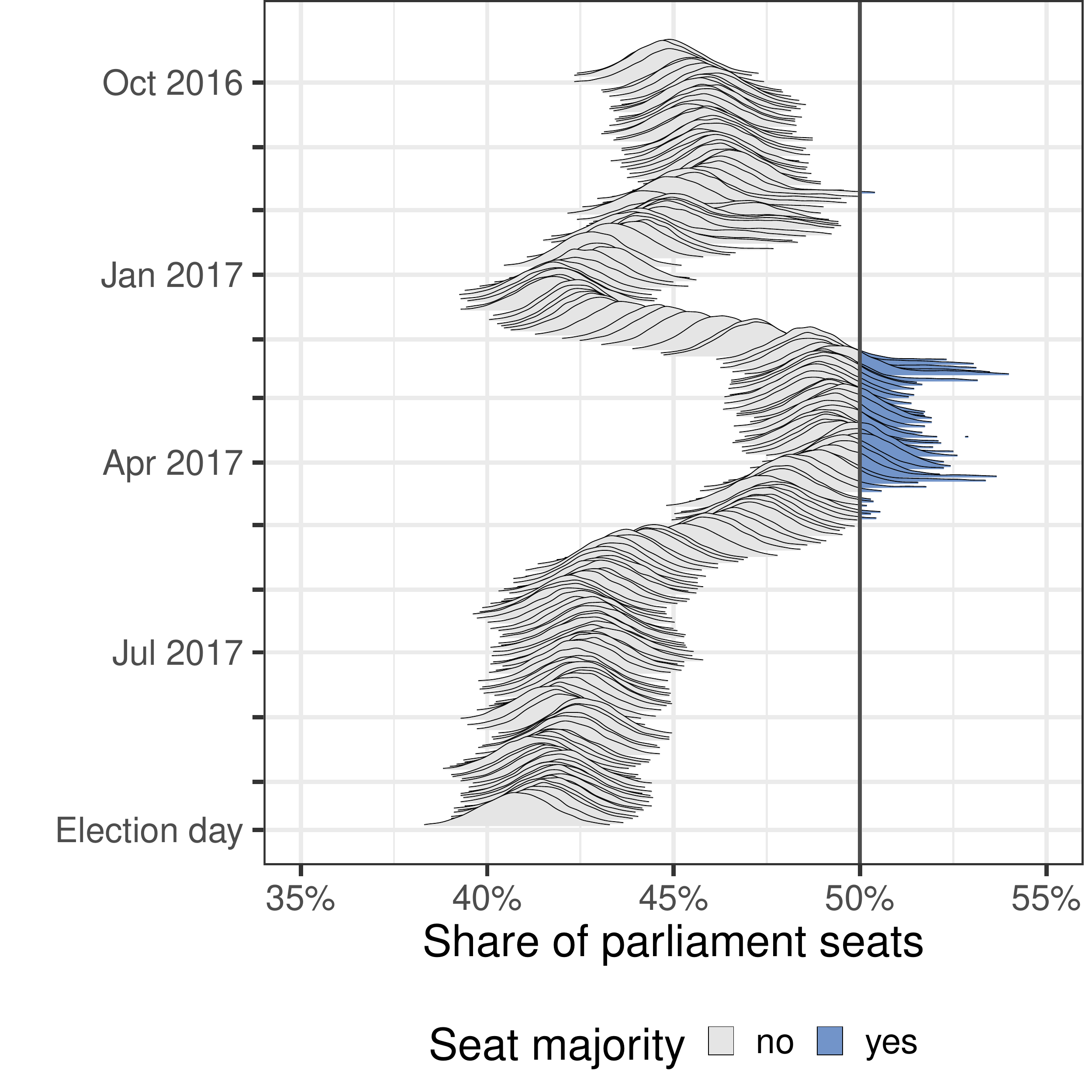}}
\\
\includegraphics[height=.2\textwidth]{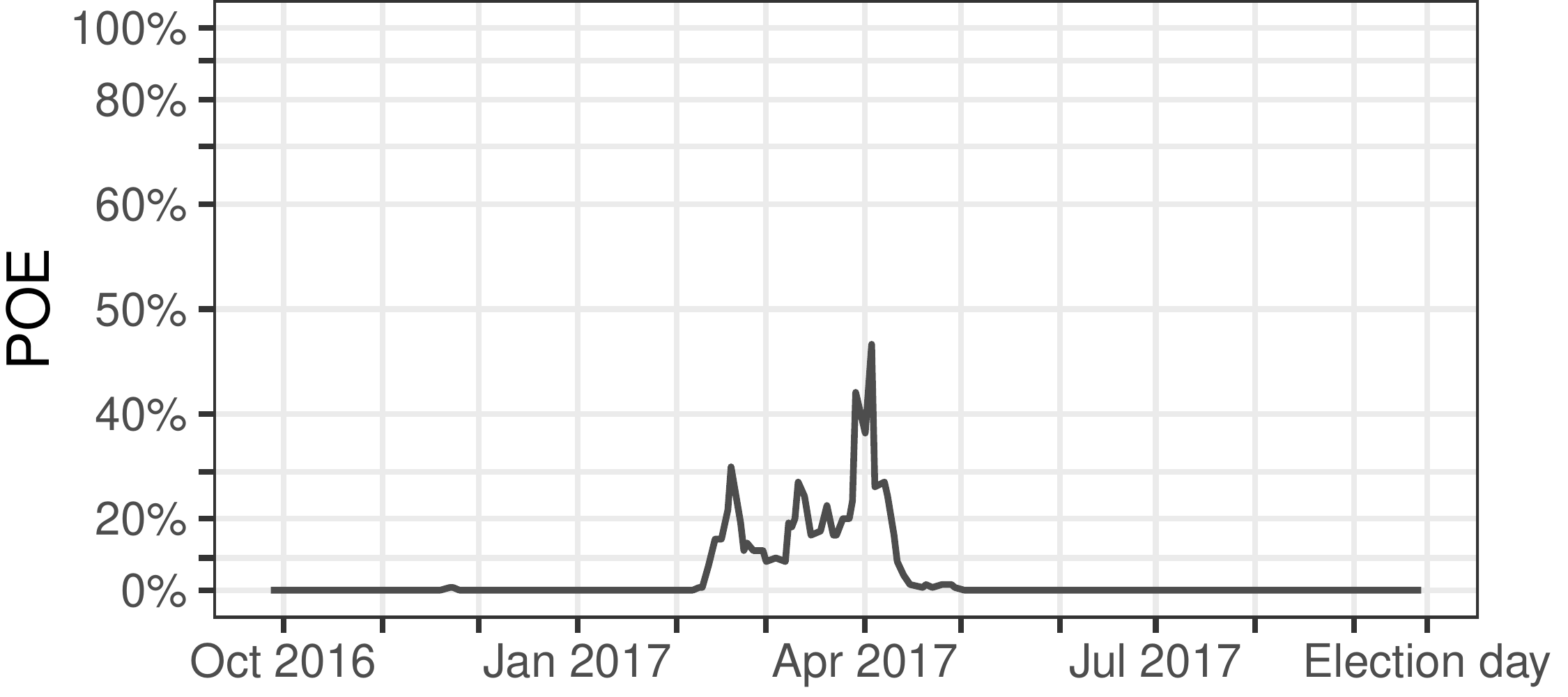}
\end{tabular}
\caption{Prospect of the coalition SPD-Left-Greens to obtain a government majority before the
German federal election in September 2017 based on pooled opinion polls.
Top left: Reported joint voter shares after redistribution.
Bottom left: Now-cast of the POE that the coalition will obtain a government
majority, based on $10\,000$ simulations.
Right: Densities of the $10\,000$ simulated parliament seat shares. Areas under
the density colored blue indicate the simulations in which the coalition
obtains a parliament seat majority.
\label{fig:2017_spdleftgreens}
}
\end{figure}

\paragraph{POE: SPD-Left-Greens coalition majority} \ \\
Regarding the party share development of the SPD, the year before
the general election in 2017 was shaped by an unusually fast increase, starting
at the end of January 2017, when Martin Schulz was elected to be the SPD chancellor
candidate and a subsequent, steady decline from April 2017 on (see Fig.~\ref{fig:2017}).
Accordingly, the coalition between SPD, the Left and the Greens had
their best joint poll results between February and May 2017 as is shown in
Figure~\ref{fig:2017_spdleftgreens}.
The maximum share was reached in April with a redistributed voter
share of $\sim 50\%$, which corresponded to a POE of obtaining the
parliament seat majority of $\sim 48\%$.
Starting in April, the POE again dropped to negligibly small values.
Shortly before election day, the joint voter share
reached a value of around $41\%$, leading to POEs of practically zero.
The ridgeline plot in Figure~\ref{fig:2017_spdleftgreens}
again nicely visualizes the uncertainty underlying the event of
interest. This is not only limited to parties forming the potential coalition,
but also includes information about all other causes of uncertainty in the data.
In November and December of 2016, for example, the seat share distribution
is clearly bimodal as in a relevant share of simulations the FDP does not
pass the $5\%$ threshold and thus more votes are redistributed to other parties
(including the SPD) in these cases.

\begin{figure}[H]\centering
\begin{tabular}{l}
\includegraphics[height=.2\textwidth]{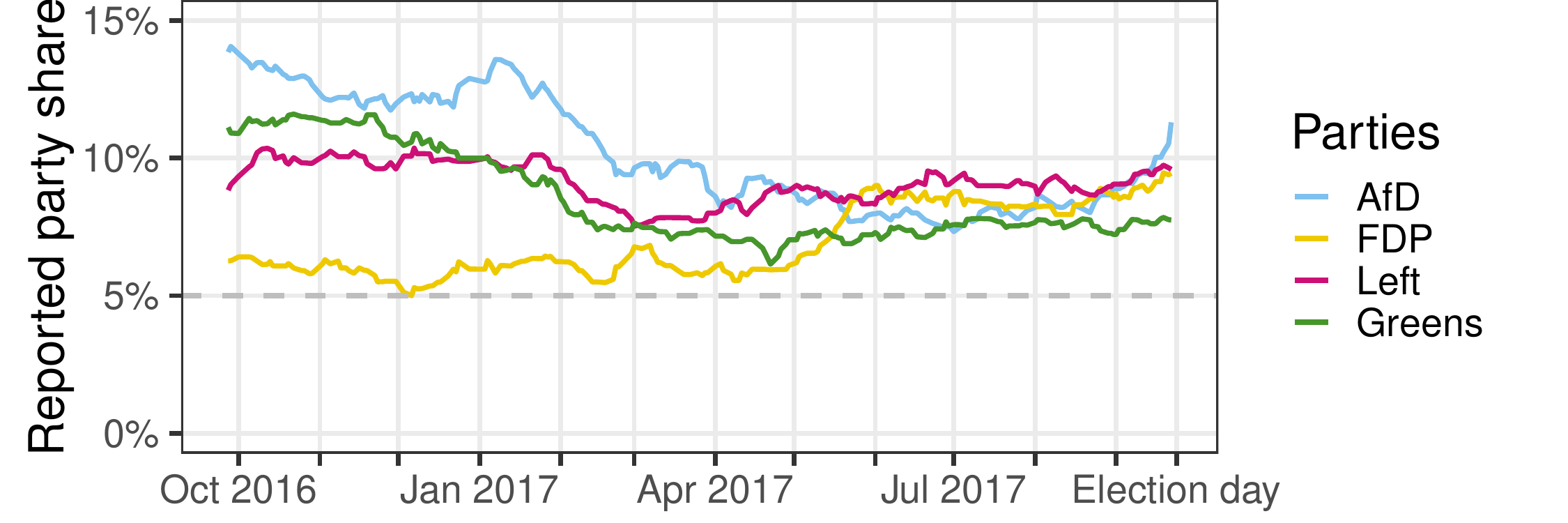}
\\
\includegraphics[height=.2\textwidth]{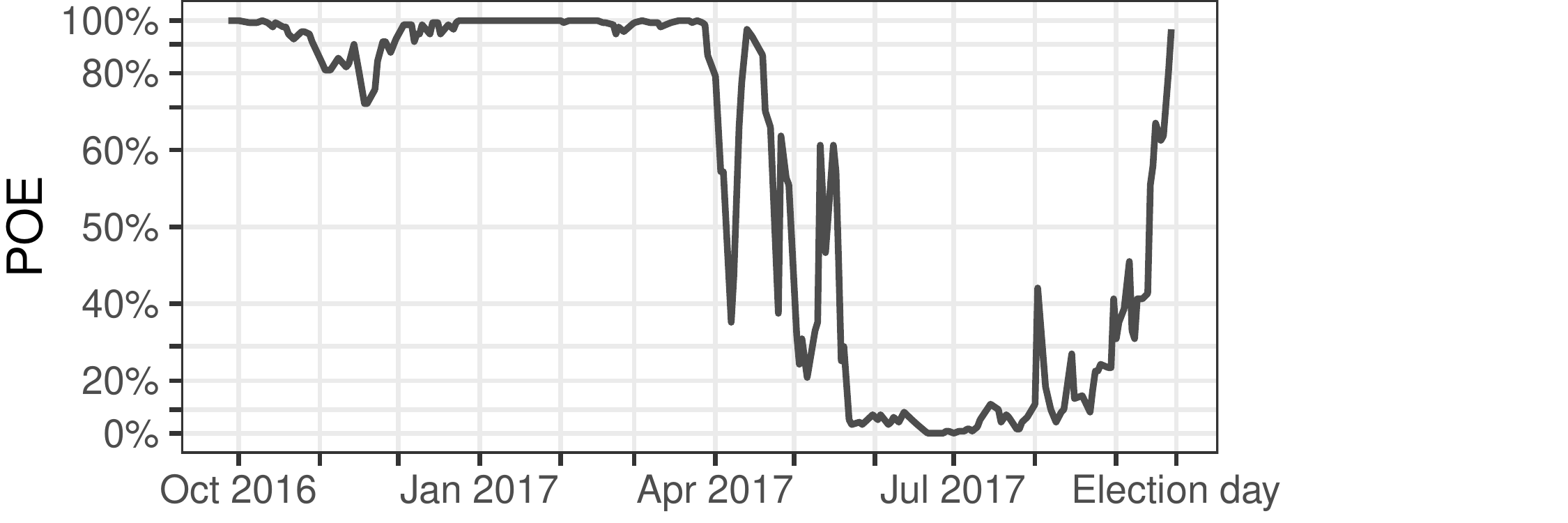}
\end{tabular}
\caption{Development of the prospect that AfD becomes the third largest party
in parliament before the German federal election in September 2017 based on pooled
opinion polls.
Top: Reported voter shares before redistribution.
Bottom: Now-cast of the POE that AfD will become the third largest party,
based on $10\,000$ simulations.
\label{fig:2017_afd}
}
\end{figure}

\paragraph{POE: AfD becoming third strongest party} \ \\
Prior to the 2017 election, special interest was on the question
which party would become the third largest party in parliament.
With reported party shares of over $8\%$, the right-wing AfD had a very good
prospect to become a member of the German parliament for the first time
(see Fig.~\ref{fig:2017_afd}) and was polling close to other smaller parties.
Using our KOALA approach, estimating the POE that the AfD becomes
the third largest party in parliament is straightforward, adequately summarizing
this event probability that simultaneously depends on all reported party shares.

In the year before the general election in 2017, reported AfD party shares
underwent strong fluctuations. In January 2017 the party had a $3.9\%$ lead
over the Left and the Greens (corresponding to an estimated POE of
becoming the third largest party in parliament of $100\%$). Subsequently, the
AfD share dropped $1.9$ percentage points behind the Left in June
(corresponding to a $1.2\%$ probability)
and rose back to a $1.7$ percentage point lead (in voter shares) over the Left
and the FDP shortly before election day (corresponding to a POE of $96.8\%$).

\section{Discussion} \label{sec:conclusion}
In this article we introduce a Bayesian approach to now-cast probabilities of
election outcome related events (POEs) based on publicly available opinion poll data
Sample uncertainty is reduced by combining polls from multiple polling agencies,
while taking into account the correlation between polls.
Rounding errors of reported party shares are also accounted for.
The estimated POEs are easy to communicate to the general public and provide
a new paradigm for election coverage and reporting of opinion polls.
More specifically, the focus on event probabilities allows to capture
changes in the current political mood and their effect on events of interest more
intuitively and comprehensively while taking into account the potentially complex
range of possible outcomes due to uncertainty inherent in the
reported party shares. The value of POE based reporting was illustrated by
application to the 2013 and 2017 German federal elections.
Various visualization techniques were used to make the result accessible to the
general public. POEs are continuously updated and
made available on a dedicated website. The methods for pre-processing and
calculation of POEs are available in the
open-source \texttt{R} package \texttt{coalitions} that allows for a straightforward
application of the method to any multi-party electoral system.\\

Our approach is based on results of opinion polls conducted by different
polling agencies. Consequently, problems with well known sources of bias induced
by non response, incorrect answers, non coverage etc. can occur. All institutes
perform some correction methods to reduce these biases, using weighting or related
procedures, but they do not make their procedures completely transparent.
Currently we perform no additional assessment or correction for potential biases
of individual polling agencies. We also do not perform forecasts or use
any other information outside of reported party shares (and sample size).\\

Our long-term goal is to make probability based reporting of opinion poll based
election coverage available to the general public. One limitation in this regard
is the availability of properly structured data and its accessibility through
an application programming interface (API). The creation of such a data base
would greatly enhance the development of our and other methods for the analysis
of opinion polls. Future iterations of our algorithms could also focus on
enhancing  the computation speed of the Monte Carlo based calculation of
POEs, making updates available to the general public even more quickly,
ideally in real time when new opinion polls are published.

%


\bibliography{bauer_2018}   

%
%

\end{document}